\documentstyle[prl,aps]{revtex}

\newcommand{\np}[3]{Nucl. Phys. {\bf B#1}, #3 (19#2)}
\renewcommand{\pl}[3]{Phys. Lett. {\bf #1B}, #3 (19#2)}
\newcommand{\pr}[3]{Phys. Rev. D {\bf #1}, #3 (19#2)}
\renewcommand{\prl}[3]{Phys. Rev. Lett. {\bf #1}, #3 (19#2)}

\newcommand{\vj}[4]{#1~{\bf #2}, #4 (19#3)}

\def\be{\begin{equation}}
\def\ee{\end{equation}}
\def\bea{\begin{eqnarray}}
\def\eea{\end{eqnarray}}

\newcommand{\nn}{\nonumber\\}

\def\ie{{\it i.e.\/}}

\newcommand{\g}{\gamma}

\newcommand{\kb}{k \cdot b}

\newcommand{\kbE}{k_E \cdot b_E}

\newcommand{\ksl}{{\not \! k}}
\newcommand{\bsl}{{\not \! b}}

\newcommand{\integ}{\int \frac{{\mathrm d^4}k}{(2\pi)^4}\,}

\begin{document}

\preprint{UAB-FT-466}          
\draft
\title{Exact calculation of the radiatively-induced \\
       Lorentz and CPT violation in QED}  
\author{M. P\'erez-Victoria}  
\address{Grup de F\'{\i}sica Te\`orica and Institut de F\'{\i}sica d'Altes
 Energies 
 (IFAE), \\  
 Universitat Aut\`onoma de Barcelona \\
 E-08193 Bellaterra, Barcelona, Spain} 
\maketitle

\begin{abstract}  
Radiative corrections arising from the axial coupling of charged
fermions to a constant vector $b_\mu$ can induce a Lorentz- and CPT-violating
Chern-Simons term in the QED action. We calculate the exact one-loop
correction to this term keeping the full $b_\mu$ dependence, and show that in
the physically interesting cases it
coincides with the lowest-order result. The effect of regularization and
renormalization and the implications of the result are briefly discussed.
\end{abstract}

\pacs{11.30.Cp, 11.30.Er, 11.30.Qc}


The possible breaking of CPT and Lorentz invariance due to non-conventional
physics has been recently addressed by constructing extensions of the standard
model that include tiny non-invariant renormalizable terms (see
Ref.\cite{CoKo1,CoKo2,CoGl,Ko_talk,Co_talk} and references therein; the
possibility of dynamical Lorentz symmetry breaking has been considered in
Ref.~\cite{AnSolPRD}). In
particular, it is 
interesting to consider the QED sector of such extensions. We shall only be
concerned here with the Lorentz-violating CPT-odd terms, which for a single
charged (Dirac) fermion read 
\be
 S^{\mbox{\tiny CPT}}= \int {\mathrm d}^4 x \left[
 -a_\mu \bar{\psi} \gamma^\mu \psi - b_\mu \bar{\psi} \gamma^\mu
 \gamma_5 \psi + \frac{1}{2} k^\mu \epsilon_{\mu\nu\rho\sigma} A^\nu
 F^{\rho\sigma} \right] \, . \label{LCPT}
\ee
Stringent experimental bounds can be  put on the pure-photon CPT-violating
term \cite{CaFiJa},  which is of the Chern-Simons form \cite{JaTe} 
(a disputed claim
exists, however, for a nonzero $\vec{k}$ \cite{NoRa,CaFi}). Moreover, this
term introduces tachyonic modes in the photon spectrum and, for a timelike
$k_\mu$, a vacuum instability ~\cite{CaFiJa,AnSolPLB}. Hence, 
both experiment and theory suggest that $k_\mu$ should vanish (at least in
the timelike case).  A natural
question is then whether a non-zero $k_\mu$ can be induced by radiative
corrections involving Lorentz and CPT-violating couplings in other sectors of
the total low-energy theory. In that case, the tight bounds on $k_\mu$ would
also constrain these sectors. In the QED extension such corrections can
only arise from the axial-vector term, with coupling $b_\mu$. 

Several authors have tried to answer this question. All calculations
have been performed to one loop and at leading
order in $b_\mu$, and have rendered a finite result. However, despite some
early claims of definite values for the induced $k_\mu$ \cite{CoGl,ChOh}, it
seems quite clear now that the result is ambiguous
\cite{CoKo2,Chung,Chen,JaKo}, \ie, 
depends on the 
details of the high-energy theory~\cite{Vo}. It is our purpose here
to calculate the
one-loop corrections to all orders in the coupling $b_\mu$ and 
discuss the relevant issues in the light of the exact result. After this work
had been completed, we learnt that J.~M. Chung had carried out the same
calculation (for $b^2<m^2$), with the same result~\cite{ChungGemelo}.

The relevant quantity is the parity-odd part of the vacuum polarization, which
must be of the form
\be
\Pi_{\mbox{\small odd}}^{\mu\nu}(p)= \epsilon^{\mu\nu\alpha\beta}b_\alpha
p_\beta K(p,b,m),
\ee
where $p_\mu$ is the external momentum and the
function $K$ is a scalar. The contribution to the induced Chern-Simons term in
the effective action is given by 
\be
(\Delta k)^\mu= - \frac{1}{2} b^\mu K(0,b,m) \,,
\ee
and must be a function of $b^2/m^2$. To one-loop, the only contributing
diagram coincides with the standard one-loop vacuum-polarization but with the
usual fermion propagator replaced by the $b_\mu$-exact propagator
\be
S_b(k)=\frac{i}{\ksl - m - \bsl \g_5} \, .
\ee
We use a hermitian $\g_5$ with 
$tr\{\g^\mu\g^\nu\g^\rho\g^\sigma\g_5\}=4i\epsilon^{\mu\nu\rho\sigma}$ and
$g_{\mu\nu}=\mbox{diag}(1,-1,-1,-1)$. In
order to keep the full  
dependence on $b_\mu$ we must rationalize the propagator. We find
\be
S_b(k)=i \frac{(\ksl +m-\bsl \g_5)(k^2-m^2-b^2+[\ksl,\bsl]
  \g_5)}{(k^2-m^2-b^2)^2 + 4 (k^2 b^2 - (\kb)^2)}\,, \label{propagator}
\ee
which agrees with the expression given in Ref.~\cite{CoKo1}.
As discussed there, this propagator has four poles that occur
at real values of $k_0$~\cite{fn1}.
The one-loop vacuum polarization reads
\be
\Pi^{\mu\nu}(p)=\integ {\mathrm tr}\left\{ \g^\mu S_b(k) \g^\nu S_b(k-p)
\right\}.
\ee
This integral is linearly divergent. Eq.~\ref{propagator} allows us to compute the trace in
the numerator, which for the odd terms in $b_\mu$ reduces to  
$\epsilon^{\mu\nu\alpha\beta} p_\beta F_\alpha(k,p,b,m)$, with 
\bea
\lefteqn{F_\alpha(k,p,b,m)= b_\alpha F^{(1)}_\alpha(k,p,b,m) + k_\alpha
  F^{(2)}_\alpha(k,p,b,m)}  &&  \nn 
&& = \mbox{} -4 i \left\{ b_\alpha \left[2m^2 (k^2-m^2-b^2)
   +(k^2+m^2+b^2)((k-p)^2-m^2+b^2)+4\kb (k-p)\cdot b \right]  \right.   \nn
&& \makebox[5mm]{} - \left. 2 k_\alpha \left[(k^2-m^2+b^2)(k-p)\cdot
    b+((k-p)^2-m^2+b^2) \kb\right]\right\} \,.
\eea
The linearly divergent term has disappeared, leaving an integral which is just
logarithmically divergent by power counting.
Since we are only interested in $K(p,b,m)$ for $p_\mu=0$ and, luckily, no term
with 
$\epsilon^{\mu\nu\alpha\beta} b_\alpha k_\beta$ appears, we can simplify the
calculation by setting $p_\mu=0$ in $F_\alpha(k,p,b,m)$ and in the denominator
of the integral. We are then left with two integrals which only depend on
$b_\mu$ and $m$. The first one is already of the form $b_\alpha
K_1(0,b,m)$ . The second integral must also give a result of the form
$b_\alpha K_2(0,b,m)$, and $K_2(0,b,m)$ can be
calculated multiplying the integral by $b_\alpha$ and dividing by $b^2$ (as
long as $b_\mu$ is not lightlike).
In this way, we arrive at the following expression:
\bea
K(0,b,m) & = & -4 i \integ \frac{1} 
  {\left(\,(k^2-m^2-b^2)^2+4(k^2 b^2-(\kb)^2)\,\right)^2} \nn
&& \left\{\rule[-.3cm]{0cm}{.6cm} \left[2m^2
    (k^2-m^2-b^2)+(k^2+b^2)^2+4(\kb)^2-m^4 
  \right] \right. \nn  
&& \left. \mbox{} - \frac{1}{b^2} \left[4 (\kb)^2 (k^2-m^2+b^2)\right]
  \rule[-.3cm]{0cm}{.6cm}  \right\} . \label{Minkint}
\eea
In order to calculate this integral, we go to
Euclidean space via a Wick rotation of $k_0$ and then perform an analytic 
continuation to ${b_E}=(ib_0,\vec{b})$: 
\bea
K(0,b_E,m) &=& - 4 \int \frac{{\mathrm d}^4 k_E}{(2\pi)^4} \, \frac{1} 
  {\left(\,(k_E^2+m^2-b_E^2)^2+4(k_E^2 b_E^2-(\kbE)^2)\,\right)^2}  \nn 
&& \left\{ \rule[-.3cm]{0cm}{.6cm} \left[2m^2
   (k_E^2+m^2-b_E^2)-(k_E^2+b_E^2)^2-4(\kbE)^2+m^4 
  \right] \right. \nn
&& \left. \mbox{} + \frac{1}{b_E^2} \left[4 (\kbE)^2 (k_E^2+m^2+b_E^2)\right]
  \rule[-.3cm]{0cm}{.6cm} \right\} \, . 
\label{eucl} 
\eea
Here the scalar product is Euclidean.
One can directly  see at this stage that the result must be finite. Indeed, for
very large  $|k_E|$ the leading term in the integrand has the form
\be
\frac{k_E^2- 4 (\kbE)^2/b_E^2}{k_E^6}, \label{leading}
\ee
which gives a vanishing result if the integral is done symmetrically. The
other terms are power-counting finite. This also shows an ambiguity in
the induced term: the integral of the expression~(\ref{leading}) is
regularization dependent! In (four-dimensional) spherical
coordinates Eq.~(\ref{eucl}) reads: 
\bea
K(0,b_E,m) &=& - \frac{1}{\pi^3} \int_0^\infty {\mathrm d}
  |k_E|\, 
  |k_E|^3 \int_0^\pi {\mathrm d}\theta\, \sin^2 \theta \frac{1}  
  {\left(\,(|k_E|^2+m^2-b_E^2)^2+4|k_E|^2 b_E^2 \sin^2\theta \,\right)^2}  \nn 
&& \left\{ \rule[-.3cm]{0cm}{.6cm} \left[2m^2
    (|k_E|^2+m^2-b_E^2)-(|k_E|^2+b_E^2)^2-4|k_E|^2b_E^2\cos^2\theta  
   +m^4 \right] \right. \nn
&& \left. \mbox{} + \left[4 |k_E|^2 (|k_E|^2+m^2+b_E^2)\cos^2\theta\right]
  \rule[-.3cm]{0cm}{.6cm}  \right\} \,. \label{intesfericas}
\eea
Doing first the angular integral we find
\bea
\lefteqn{K(0,b_E,m) = \frac{1}{4 \pi^2 b_E^4} \int_0^\infty {\mathrm d}
  |k_E|\, |k_E| \left\{ \rule[-.4cm]{0cm}{.8cm} (k^2+m^2) -
    \mbox{Sign}\left(|k_E|^2+m^2-b_E^2\right) 
  \right.} && \nn 
&& \left. 
  \frac{(|k_E|^2+m^2)^4+3b_E^2(|k_E|^2+m^2)^2(|k_E|^2+b_E^2-m^2)+b_E^6(|k_E|^2-m^2)}
  {\left(4b_E^2|k_E|^2 + (|k_E|^2+m^2-b_E^2)^2 \right)^{3/2}}
  \rule[-.4cm]{0cm}{.8cm} \right\} 
  \, .  
\eea
This integral is well-behaved for large $|k_E|$. Note the appearance of the
sign function. For $m\not=0$, the final result is (going back to the
Minkowskian $b_\mu$): 
\bea
&&\bullet ~~~ (\Delta k)^\mu = \frac{3}{16\pi^2} b^\mu \,  ,  
  \mbox{~~~if $-b^2 \leq m^2$} ~; \label{timesol}
\\  
&&\nn
&&\bullet ~~~ (\Delta k)^\mu = \left(\frac{3}{16\pi^2} - \frac{1}{4\pi^2}
                         \sqrt{1-\frac{m^2}{|b^2|}}\right) b^\mu \, 
   ,  \mbox{~~~if $-b^2 > m^2$} ~. \label{spacesol}
\eea
For any timelike $b_\mu$ and for a spacelike $b_\mu$ with $|b^2| < m^2$,
Eq.~(\ref{timesol}) is the relevant one. Surprisingly enough, in these cases
our calculation to all orders in $b_\mu$  gives the same result as the
one obtained in the $b_\mu$-linear approximation of Ref.~\cite{JaKo}.
Obviously, perturbation theory about $b_\mu=0$ does not detect the different
behaviour we have found for $-b^2 > m^2$. On the other hand, continuity in
$b^2$ implies that the lightlike case, $b^2=0$, is also given by
Eq.~(\ref{timesol}). In fact, dimensional analysis shows that the
$b_\mu$-linear
approximation is exact for vanishing $b^2$. Note also in passing that for the
fine-tuned value $b^2=-\frac{16}{7} m^2$, a vanishing $(\Delta k)^\mu$ is
obtained.

If the fermion is massless, $m=0$, we find $(\Delta k)^\mu =
-\frac{1}{16\pi^2} b^\mu$ for any kind of $b_\mu$. In this simple case,
there are at least three other possible  ways
of calculating the induced term, which give the same
answer: 
\begin{enumerate}
\item A simpler $b_\mu$-exact propagator can obtained for $m=0$:
\be
S_b(k)=i \frac{(\ksl +\bsl \g_5)(k^2+b^2-2 \kb \g_5)}
  {(k+b)^2(k-b)^2}\,.
\ee
The calculation can then be simplified using Feynman parameters and the result
$(\Delta k)^\mu = -\frac{1}{16\pi^2} b^\mu$ is found for
any $b_\mu$. 
\item Perturbatively, one can perform just a $b_\mu$-linear calculation, since
  in the massless case the higher-order terms vanish for dimensional
  reasons. From Eq.~(14) in Ref.~\cite{JaKo}, the same $(\Delta k)^\mu$
  results. The only contribution comes from the surface term in Eq.~(10)
  of Jackiw-Kosteleck\'y's calculation.
\item A confirmation of the same result, non-perturbative in $b_\mu$ and in the
  fine-structure constant,
  is provided in Ref.~\cite{Chung} (following a suggestion by D. Colladay): an
  anomalous chiral redefinition of 
  the fermion fields allows to get rid of the coupling to $b_\mu$, so that the
  contribution (to all orders) to $K(0,b,0)$ comes from the corresponding
  Fujikawa Jacobian. Up to the unavoidable ambiguity (which in this method
  comes from the definition of the current operator), 
  $-\frac{1}{16\pi^2} b^\mu$ is obtained again. 
\end{enumerate}
An infrared regularization can spoil
this result. For instance, giving the fermion a small mass obviously  shifts
it back to $\frac{3}{16\pi^2} b^\mu$ in 
the timelike-$b_\mu$ case.
At any rate this is just a formal discussion, since there are no massless
electromagnetically-charged fermions in nature. 

We have also calculated the integral in Eq.~\ref{Minkint} directly in
Minkowski space,  using first the residues
method to perform the integration on $k_0$, and integrating on $\vec{k}$
afterwards. Although the result differs by a
constant, because $k_0$ and $\vec{k}$ are not treated symmetrically,
the same dependence on $b^2$ and $m^2$ is found.

It is rather striking that the contributions to $(\Delta k)^\mu$ of
diagrams with more than one insertion of $\bsl\g_5$ vanish. We have explicitly
checked that this is indeed the case at order $b_\mu b^2$. At this and higher
orders, all integrals are finite by power-counting, and hence
unambigous. S. Coleman has observed that the vanishing of these higher order
contributions can be explained by his argument with S.L. Glashow in
Ref.~\cite{CoGl}, which can be applied to finite diagrams with insertions
of $\bsl\g_5$~\cite{fn2}. Consider a two-photon amplitude with $n>1$
insertions of $\bsl\g_5$. The idea is to let each of the two photons carry
different momenta, $p$ and $q$ (the insertions carry non-zero
momentum then). Gauge invariance implies transversality in each of the
photons. Differentiating each of the transversality conditions with respect
to the corresponding momentum, one learns that the amplitude is $O(pq)$.
It follows that when ones goes to equal momenta, $q=p$, the amplitude is
$O(p^2)$. Since 
the Chern-Simons term in the effective action is $O(p)$,  one concludes that
this amplitude does not contribute to $(\Delta k)^\mu$. In Ref.~\cite{CoGl},
this argument is proved to be valid to any order in the
fine-structure constant. Note, however, that it
does not apply to diagrams with just one insertion, due to the presence of
triangular anomalies~\cite{JaKo}.

Let us discuss now how regularization and renormalization affect the
result. This is an important point because the complete 
$S^{\mbox{\tiny QED}} + S^{\mbox{\tiny CPT}}$ theory is not
finite and requires renormalization (and, furthermore, renormalization is
also relevant in a finite theory~\cite{Weinberg,Jatalk}). The {\em exact}
decomposition $S_b(k)=S(k)-iS_b(k)\bsl\g_5S(k)$ performed in Ref.~\cite{JaKo},
shows that the ambiguities can only come from the lowest-order piece, the rest
being finite by power counting.
In our calculation, the result can be 
changed by any regularization that destroys either the spherical
symmetry (in four-dimensional Euclidean space) of the high-energy
behaviour or the steps we followed to arrive at Eq.~(\ref{eucl}). In general,
different regularizations (or subtractions) will render different
results, even if they preserve gauge invariance. This is apparent in
differential renormalization, which makes the ambiguity
explicit~\cite{Chen}.
As a matter of fact, independently of how one regulates and subtracts the
divergent integrals, one always has the freedom to add any
(renormalizable) finite counterterm 
that is allowed by the relevant symmetries of the theory~\cite{PiSo}. This is
also true when the radiative corrections to that term are
finite~\cite{Jatalk}. In the present case, this means that the induced $(\Delta
k)^\mu$ can have any value, for it is not protected by any
symmetry~\cite{JaKo} (except CPT and Lorentz invariance, but we just broke
them). The conclusion of our study is the following: if the regulator
and the subtraction rule are mass-independent, a mass-independent result will
be obtained to all orders in $b_\mu$ in the physically-relevant cases, as the
CPT- and Lorentz-violating terms coefficients are 
much smaller than the mass of any electromagnetically-charged fermion and
Eq.~(\ref{timesol}) 
provides the induced term in this situation.

In Ref.~\cite{CoKo2} an interesting discussion was made regarding the possible
vanishing of the induced Chern-Simons term due to an anomaly-cancellation
mechanism in the high-energy theory (of course, we consider now several
fermion species). Essentially, the argument goes as
follows. From the point of view of a more fundamental theory, the
diagrams with one$\bsl\g_5$ insertion (at one loop) can be viewed as the
corresponding triangular diagrams with the same photon legs and a third leg
involving a coupling to an axial vector, in the limit in which there is
zero momentum transfer to the axial-vector leg and the latter is replaced with
a vacuum expectation value. The condition for the cancellation of the anomalies
ocurring in these diagrams then implies that the induced term also cancels. 
This argument requires that the term induced by different fermions be the
same. This is true if the induced term contains no mass dependence and,
besides, a universal and mass-independent renormalization prescription is
adopted for all the contributing diagrams (such a prescription can be
justified again by a similar argument). Our result 
shows that the first 
requirement, which is trivial at the $b_\mu$-linear order, holds to all orders
in $b_\mu$. Invoking the Adler-Bardeen
theorem, the argument for the vanishing of the
induced term was also generalized in Ref.~\cite{CoKo2} to higher loops. 
Again, this is only valid if no
mass-dependence is introduced by higher-order corrections. That this is the
case follows from the combination of Coleman-Glashow's
argument and Adler-Bardeen theorem. Summarizing: in the context of
Ref.~\cite{CoKo2}, the cancellation of anomalies
in the fundamental theory imply the vanishing of the induced term to 
any order in $b_\mu$ and in the fine-structure constant, if a mass-independent
scheme is used.

We have till now neglected the
possible influence of the $a_\mu$ term in Eq.~(\ref{LCPT}) but, in principle,
corrections of order $a^2 b_\mu$ and higher could appear. Actually, one can
include the effect of $a_\mu$ to all orders by considering the
corresponding $a_\mu$- 
and $b_\mu$-exact propagator. This propagator is just the one in
Eq.~(\ref{propagator}) but with $k_\mu$ substituted by $k_\mu-a_\mu$. The
vector $a_\mu$
behaves then like an external momentum which appears in all the propagators of
the loop. Since there are no derivative couplings, a shift in the loop
momentum $k_\mu\rightarrow k_\mu+a_\mu$  can completely eliminate $a_\mu$, so
the result is not affected. This shift is also subjected to
regularization ambiguities. As a matter of fact, the term proportional to
$a_\mu$ can be eliminated from the action
$S^{\mbox{\tiny QED}} + S^{\mbox{\tiny CPT}}$ by a field
redefinition of the form $\psi=\exp(-i a\cdot x) \chi$~\cite{CoKo1}.  
The effect of other sectors (like the CPT-even Lorentz-violating
extension of QED considered in Ref.~\cite{CoKo2}) can also be studied with
these techniques, \ie, incorporating
the corresponding fermion bilinears into the exact propagator. This is beyond
the scope of the present work.

Let us finally stress that even if the radiative corrections to the
Chern-Simons term cancel for some reason, it is still possible to add a finite
counterterm and get a non-zero $(\Delta k)^\mu$. This is a sign of the fact
that we have no right to put $k^\mu=0$ at tree level, unless there is some
symmetry in the high-energy theory that imposes this value. Of course, it is
comparison with experiment that tells us to set $k^\mu+(\Delta k)^\mu=0$ but,
in the absence of a proper symmetry, we are facing a problem of naturalness.
Nevertheless, the anomaly-cancellation
argument shows that this problem does not come from other sectors of the
low-energy theory, and, from a practical point of view, allows one to put
$k^\mu=0$ in the tree-level action,  as
long as a consistent mass-independent scheme is used in the calculations.

\acknowledgments
We thank F. del Aguila, F. Ferrer and S. Peris for discussions, and
D. Colladay, R. Jackiw and V.~A. Kosteleck\'y for many useful
comments. This work has been partially supported by CICYT, AEN96-1672 and
Junta de Andaluc\'{\i}a, FQM101.

\end{document}